\PassOptionsToPackage{unicode}{hyperref}
\PassOptionsToPackage{hyphens}{url}
\PassOptionsToPackage{dvipsnames,svgnames,x11names}{xcolor}
\documentclass[
  10pt,
  runningheads]{llncs}
\usepackage{xcolor}
\usepackage{amsmath,amssymb}
\usepackage{iftex}
\ifPDFTeX
  \usepackage[T1]{fontenc}
  \usepackage[utf8]{inputenc}
  \usepackage{textcomp} 
\else 
  \usepackage{unicode-math} 
  \defaultfontfeatures{Scale=MatchLowercase}
  \defaultfontfeatures[\rmfamily]{Ligatures=TeX,Scale=1}
\fi
\usepackage{lmodern}
\ifPDFTeX\else
\fi
\IfFileExists{upquote.sty}{\usepackage{upquote}}{}
\IfFileExists{microtype.sty}{
  \usepackage[]{microtype}
  \UseMicrotypeSet[protrusion]{basicmath} 
}{}
\makeatletter
\@ifundefined{KOMAClassName}{
  \IfFileExists{parskip.sty}{%
    \usepackage{parskip}
  }{
    \setlength{\parindent}{0pt}
    \setlength{\parskip}{6pt plus 2pt minus 1pt}}
}{
  \KOMAoptions{parskip=half}}
\makeatother
\makeatletter
\ifx\paragraph\undefined\else
  \let\oldparagraph\paragraph
  \renewcommand{\paragraph}{
    \@ifstar
      \xxxParagraphStar
      \xxxParagraphNoStar
  }
  \newcommand{\xxxParagraphStar}[1]{\oldparagraph*{#1}\mbox{}}
  \newcommand{\xxxParagraphNoStar}[1]{\oldparagraph{#1}\mbox{}}
\fi
\ifx\subparagraph\undefined\else
  \let\oldsubparagraph\subparagraph
  \renewcommand{\subparagraph}{
    \@ifstar
      \xxxSubParagraphStar
      \xxxSubParagraphNoStar
  }
  \newcommand{\xxxSubParagraphStar}[1]{\oldsubparagraph*{#1}\mbox{}}
  \newcommand{\xxxSubParagraphNoStar}[1]{\oldsubparagraph{#1}\mbox{}}
\fi
\makeatother

\usepackage{longtable,booktabs,array}
\usepackage{calc} 
\usepackage{etoolbox}
\makeatletter
\patchcmd\longtable{\par}{\if@noskipsec\mbox{}\fi\par}{}{}
\makeatother
\IfFileExists{footnotehyper.sty}{\usepackage{footnotehyper}}{\usepackage{footnote}}
\makesavenoteenv{longtable}
\usepackage{graphicx}
\makeatletter
\newsavebox\pandoc@box
\newcommand*\pandocbounded[1]{
  \sbox\pandoc@box{#1}%
  \Gscale@div\@tempa{\textheight}{\dimexpr\ht\pandoc@box+\dp\pandoc@box\relax}%
  \Gscale@div\@tempb{\linewidth}{\wd\pandoc@box}%
  \ifdim\@tempb\p@<\@tempa\p@\let\@tempa\@tempb\fi
  \ifdim\@tempa\p@<\p@\scalebox{\@tempa}{\usebox\pandoc@box}%
  \else\usebox{\pandoc@box}%
  \fi%
}
\def\fps@figure{htbp}
\makeatother

\providecommand{\tightlist}{%
  \setlength{\itemsep}{0pt}\setlength{\parskip}{0pt}}

\usepackage[numbers]{natbib}
\titlerunning{Agent Contracts: Resource-Bounded Autonomous AI}
\authorrunning{Q. Ye and J. Tan}
\AtBeginDocument{%
  \author{Qing Ye\inst{1} \and Jing Tan\inst{2}}%
  \institute{Independent Researcher \email{yeqi519@gmail.com} \and Independent Researcher \email{jtan@live.de}}%
}
\makeatletter
\@ifpackageloaded{caption}{}{\usepackage{caption}}
\AtBeginDocument{%
\ifdefined\contentsname
  \renewcommand*\contentsname{Table of contents}
\else
  \newcommand\contentsname{Table of contents}
\fi
\ifdefined\listfigurename
  \renewcommand*\listfigurename{List of Figures}
\else
  \newcommand\listfigurename{List of Figures}
\fi
\ifdefined\listtablename
  \renewcommand*\listtablename{List of Tables}
\else
  \newcommand\listtablename{List of Tables}
\fi
\ifdefined\figurename
  \renewcommand*\figurename{Figure}
\else
  \newcommand\figurename{Figure}
\fi
\ifdefined\tablename
  \renewcommand*\tablename{Table}
\else
  \newcommand\tablename{Table}
\fi
}
\@ifpackageloaded{float}{}{\usepackage{float}}
\floatstyle{ruled}
\@ifundefined{c@chapter}{\newfloat{codelisting}{h}{lop}}{\newfloat{codelisting}{h}{lop}[chapter]}
\floatname{codelisting}{Listing}

\makeatother
\makeatletter
\@ifpackageloaded{caption}{}{\usepackage{caption}}
\@ifpackageloaded{subcaption}{}{\usepackage{subcaption}}
\makeatother
\usepackage{bookmark}
\IfFileExists{xurl.sty}{\usepackage{xurl}}{} 
\hypersetup{
  pdftitle={Agent Contracts: A Formal Framework for Resource-Bounded Autonomous AI Systems (Full)},
  pdfauthor={Qing Ye; Jing Tan},
  pdfkeywords={AI agents, resource governance, multi-agent
systems, normative constraints, contract theory, agent coordination},
  colorlinks=true,
  linkcolor={blue},
  filecolor={Maroon},
  citecolor={Blue},
  urlcolor={Blue},
  pdfcreator={LaTeX via pandoc}}

\title{Agent Contracts: A Formal Framework for Resource-Bounded
Autonomous AI Systems
(Full)\thanks{Accepted for oral presentation at COINE 2026 (16th International Workshop on Coordination, Organizations, Institutions, Norms and Ethics for Governance of Multi-Agent Systems), co-located with AAMAS 2026, Paphos, Cyprus.}}
\author{Qing Ye \and Jing Tan}
\date{}
\begin{document}
\maketitle
\begin{abstract}
The Contract Net Protocol (1980) introduced coordination through
contracts in multi-agent systems. Modern agent protocols standardize
connectivity and interoperability, yet none provide formal resource
governance---normative mechanisms to bound \emph{how much} agents may
consume or \emph{how long} they may operate. We introduce \emph{Agent
Contracts}, a formal framework that extends the contract metaphor from
task allocation to resource-bounded execution. An Agent Contract
\(C = (I, O, S, R, T,
\Phi, \Psi)\) unifies input/output specifications, multi-dimensional
resource constraints, temporal boundaries, and success criteria into a
coherent governance mechanism with explicit lifecycle semantics. For
multi-agent coordination, we establish conservation laws ensuring
delegated budgets respect parent constraints, enabling hierarchical
coordination through contract delegation. Empirical validation across
four experiments demonstrates 90\% token reduction with 525× lower
variance in iterative workflows, zero conservation violations in
multi-agent delegation, and measurable quality-resource tradeoffs
through contract modes. Agent Contracts provide formal foundations for
predictable, auditable, and resource-bounded autonomous AI deployment.
\end{abstract}

\section{Introduction}\label{introduction}

In late 2025, an engineering team deployed a multi-agent research system
with four specialized agents. Two agents fell into a recursive
clarification loop, running undetected for eleven days. When the invoice
arrived, the team discovered a \$47,000 API bill
\citep{kusireddy2025agent}. The system had no stop conditions, no budget
limits, and no real-time cost monitoring. This incident encapsulates a
fundamental problem---we have built AI agents capable of autonomous
action but lack formal mechanisms to bound their behavior.

Such failures reflect systemic gaps, not implementation bugs
\citep{cemri2025multiagent}. Gartner predicts that over 40\% of agentic
AI projects will be canceled by 2027 due to escalating costs or
inadequate risk controls \citep{gartner2025cancel}, even as agentic AI
is projected to appear in 33\% of enterprise software by 2028
\citep{gartner2024trends}. A recent MIT Sloan study finds that 35\% of
organizations already deploy agentic AI, with leaders citing the tension
between \emph{supervision and autonomy} as a core challenge requiring
centralized governance infrastructure \citep{mitsloan2025agentic}.
Agents are becoming capable of sustained autonomous operation spanning
hours or days \citep{openai2025deepresearch, anthropic2025sonnet45}, yet
the protocols governing them address \emph{connectivity} and
\emph{interoperability} but not \emph{resource governance}---how much an
agent may consume or how long it may operate.

\textbf{Agent Contracts} address this gap by extending the contract
metaphor from task allocation to resource governance. Where the Contract
Net Protocol \citep{smith1980contract} asks ``who should do this
task?'', Agent Contracts ask ``within what bounds may this task be
performed?'' We draw on contract theory from economics, coordination
theory from distributed systems, and resource-bounded computation from
real-time systems to form a formal framework for resource-bounded
autonomous AI.

This paper makes two contributions. First, we define an Agent Contract
as a formal tuple \(C = (I, O, S, R, T, \Phi, \Psi)\) that unifies
input/output specifications, resource constraints, temporal boundaries,
and success criteria into a coherent governance mechanism. Second, we
establish conservation laws for multi-agent systems that ensure budget
discipline across delegation hierarchies, enabling composable
coordination patterns where contracting itself becomes an agent
capability. We validate these contributions across four experiments
demonstrating 90\% token reduction in iterative workflows, zero
conservation violations in multi-agent delegation, and measurable
quality-resource tradeoffs. Together, these contributions provide formal
foundations for explicit resource governance in autonomous AI systems.

\section{Theoretical Foundations}\label{theoretical-foundations}

Agent Contracts draw on three theoretical traditions: contract theory
from economics, multi-agent coordination from computer science, and
resource-bounded computation from real-time systems.

\textbf{Contract Theory.} Bolton and Dewatripont (2005)
\citep{bolton2005contract} formalize how agreements are constructed
under asymmetric information. Three concepts apply to agent governance:
\emph{moral hazard} (hidden actions; in LLM agents, unpredictable
resource consumption), \emph{incomplete contracts} (separating success
criteria from execution strategies) \citep{hart1995firms}, and
\emph{mechanism design} (specifications that elicit desired behavior).
Recent work \citep{phelps2023models, kaltenpoth2025atlas} identifies
principal-agent dynamics in LLM systems but provides conceptual rather
than operational frameworks.

\textbf{Multi-Agent Coordination.} Classical MAS research
\citep{wooldridge2009introduction, shoham2008multiagent} established
foundations for agent coordination. The Contract Net Protocol
\citep{smith1980contract} demonstrated that explicit contracting could
coordinate distributed systems. Coordination theory
\citep{malone1994interdisciplinary} identifies fundamental problems
(managing shared resources, producer-consumer relationships, and
simultaneity constraints), each corresponding to challenges in
multi-agent LLM systems. Research on normative multi-agent systems
\citep{boella2006introduction} formalizes how norms govern agent
behavior through obligations, prohibitions, and permissions. Closely
related, social commitments
\citep{singh1999commitments, castelfranchi2000commitment} endow agents
with normative accountability: a committed agent that violates its
obligation can be sanctioned. Agent Contracts operationalize the
normative perspective for LLM systems: resource constraints function as
prohibitions (agents \emph{must not} exceed budgets), success criteria
as obligations (agents \emph{must} achieve quality thresholds), and the
contract lifecycle provides regimented enforcement where violations
trigger automatic termination. However, unlike classical committed
agents, LLM agents are typically ephemeral---instantiated per task and
discarded---making agent-level sanctions inapplicable. We address this
asymmetry through structural enforcement (Section 7.2).

A key insight from MAS theory is that coordination mechanisms must
respect \emph{conservation laws}---resources allocated to subtasks
cannot exceed parent resources. This principle requires explicit
enforcement in LLM systems where token consumption is stochastic and
observable only after the fact.

\textbf{Resource-Bounded Computation.} Simon's theory of bounded
rationality \citep{simon1955behavioral, simon1956rational} established
that agents with limited cognitive resources must \emph{satisfice}
rather than maximize. Agent Contracts operationalize satisficing by
defining acceptable quality thresholds within resource budgets.

The algorithmic foundation comes from contract algorithms
\citep{zilberstein1995contract}, which specify computation budgets
\emph{before} activation. Unlike anytime algorithms
\citep{zilberstein1996anytime} that can be interrupted arbitrarily,
contract algorithms enable strategic resource allocation. An Agent
Contract transforms an LLM agent into a contract algorithm where bounds
\(R\) are known in advance. Real-time systems theory
\citep{buttazzo2011hard} contributes the distinction between hard
constraints (violation causes termination) and soft constraints
(permitting graceful degradation).

\section{Related Work}\label{related-work}

\subsection{Agent Architectures and Coordination
Protocols}\label{agent-architectures-and-coordination-protocols}

The development of LLM-based agents has accelerated rapidly since 2022.
ReAct \citep{yao2023react} introduced the paradigm of synergizing
reasoning and acting. Chain-of-Thought prompting \citep{wei2022chain}
established that computational depth correlates with output quality,
motivating explicit resource governance. Toolformer
\citep{schick2023toolformer} showed that language models can learn to
use external tools, expanding the resource consumption profile of
agents. Fully autonomous systems like AutoGPT
\citep{significant2023autogpt} and Generative Agents
\citep{park2023generative} revealed both the potential and governance
challenges of unbounded execution.

Coordination protocols have evolved to address different concerns. The
Contract Net Protocol \citep{smith1980contract} established task
allocation through bidding; MCP \citep{anthropic2024mcp} standardizes
tool connectivity between models and external resources; A2A
\citep{google2025a2a} enables discovery and interoperability across
heterogeneous agent systems. The recent formation of the Agentic AI
Foundation \citep{aaif2025} under the Linux Foundation---with
contributions including MCP, OpenAI's AGENTS.md, and Block's
goose---signals industry consensus on connectivity and documentation
standards. However, resource governance remains outside this scope; none
of these initiatives formalize how much agents may consume.

\subsection{Budget-Aware Reasoning and Resource
Management}\label{budget-aware-reasoning-and-resource-management}

A growing body of work addresses resource efficiency in LLM reasoning.
The TALE framework \citep{han2024tokenbudget} introduces
token-budget-aware reasoning, achieving 68\% reduction in token usage
with less than 5\% accuracy degradation. Critically, the authors
identify ``token elasticity'': LLMs often exceed specified budgets when
constraints are tight, demonstrating that prompting alone is
insufficient for strict enforcement. BudgetThinker
\citep{liu2025budgetthinker} addresses this through control tokens
injected during inference, coupled with reinforcement learning to
achieve precise budget adherence. SelfBudgeter
\citep{wang2025selfbudgeter} enables models to predict required token
budgets based on task complexity. Liu et al.~(2025) \citep{liu2025bats}
demonstrate that simply granting larger tool-call budgets fails to
improve agent performance; their Budget-Aware Tool Selection (BATS)
framework shows that explicit budget awareness enables effective
scaling.

Snell et al.~(2024) \citep{wang2024reasoning} provide a budget-aware
evaluation framework, demonstrating that when compute is equalized,
sophisticated reasoning strategies often do not outperform simpler
baselines; much apparent improvement comes from using more resources
rather than using them more intelligently. This finding underscores the
importance of explicit resource accounting.

Infrastructure-level resource management has matured separately. LLM
serving systems \citep{yu2022orca, kwon2023efficient} optimize
throughput and memory efficiency at the inference layer. LLMOps
platforms provide budget tracking, alerting, and rate limiting at the
organizational level. However, these operate below or above the
application layer; neither provides formal contracts that govern
individual agent behavior within multi-agent workflows.

\subsection{Agent Safety and Formal
Verification}\label{agent-safety-and-formal-verification}

Foundational AI safety work \citep{amodei2016concrete, russell2019human}
identifies problems including reward hacking and safe exploration.
Recent work calls for responsible LLM-empowered multi-agent systems
\citep{hu2025responsible}, recognizing that uncertainties compound
across agent interactions.

Formal verification approaches have begun addressing agentic AI
specifically. Zhang et al.~(2025) \citep{zhang2025formalizing} propose a
modeling framework with 17 properties for host agents and 14 for task
lifecycles, expressed in temporal logic. This work is complementary to
Agent Contracts. Formal verification addresses \emph{whether} systems
satisfy properties; Agent Contracts specify \emph{what} resource
constraints systems must satisfy.

\subsection{Multi-Agent Coordination
Frameworks}\label{multi-agent-coordination-frameworks}

LLM-based multi-agent systems have proliferated with frameworks
addressing different coordination paradigms. MetaGPT
\citep{hong2023metagpt} integrates human workflow patterns into
multi-agent collaboration, using standardized operating procedures to
structure agent interactions. AutoGen \citep{wu2023autogen} frames
coordination as asynchronous conversation among specialized agents, with
each agent capable of responding, reflecting, or invoking tools based on
message content. LangGraph \citep{langgraph2024} provides graph-based
orchestration with explicit state management and checkpointing, enabling
durable execution of complex workflows. CrewAI \citep{crewai2024} takes
a role-based approach where agents are assigned organizational roles
(researcher, developer, etc.) with corresponding capabilities.

Recent surveys characterize the landscape systematically. Tran et
al.~(2025) \citep{tran2025survey} analyze collaboration mechanisms
across five dimensions: actors involved, interaction types (cooperative,
competitive, or coopetitive), organizational structures, coordination
strategies, and communication protocols. Additional surveys examine
architectural patterns
\citep{masterman2024landscape, derouiche2025agentic} and how
message-passing architectures affect coordination effectiveness
\citep{wang2025communication}. These taxonomies reveal sophisticated
pattern vocabulary but consistently note that resource governance
remains underdeveloped. Research on self-resource allocation
\citep{li2025selfresource} demonstrates that LLMs can serve as effective
resource allocators, with planner-based approaches outperforming
real-time orchestration for concurrent task management. However, this
work studies allocation \emph{capability} rather than providing a formal
governance \emph{framework}.

To clarify this governance gap, Table~\ref{tbl-framework-comparison}
summarizes resource governance features across eight major agent
frameworks. All provide operational controls (iteration limits,
timeouts, and rate limiting), reflecting engineering best practices for
preventing runaway execution. However, none provide the formal
governance layer that Agent Contracts introduce: cost budgets, temporal
deadlines, success criteria, or conservation laws for multi-agent
delegation.

\begin{longtable}[]{@{}lcccccccc@{}}
\caption{Governance features across agent frameworks. Y = native, P =
partial, -- = none. \emph{Italic}: operational controls. \textbf{Bold}:
governance features unique to Agent Contracts. LG = LangGraph, AG =
AutoGen, Crew = CrewAI, OAI = OpenAI Agents SDK, ADK = Google ADK, BR =
Amazon Bedrock, LI = LlamaIndex, smol = smolagents.
\textsuperscript{a}AutoGen is being merged with Semantic Kernel into
Microsoft Agent Framework.
\citep{langgraph2025docs, autogen2025docs, crewai2025docs, openai2025agentssdk, google2025adk, aws2025bedrock, llamaindex2025docs, huggingface2025smolagents}}\label{tbl-framework-comparison}\tabularnewline
\toprule\noalign{}
& LG & AG\textsuperscript{a} & Crew & OAI & ADK & BR & LI & smol \\
\midrule\noalign{}
\endfirsthead
\toprule\noalign{}
& LG & AG\textsuperscript{a} & Crew & OAI & ADK & BR & LI & smol \\
\midrule\noalign{}
\endhead
\bottomrule\noalign{}
\endlastfoot
\emph{Max iterations} & Y & Y & Y & Y & Y & Y & Y & Y \\
\emph{Timeout} & Y & Y & Y & Y & Y & Y & Y & Y \\
\emph{Rate limiting} & Y & P & Y & P & P & Y & P & Y \\
\emph{Token limits} & P & Y & Y & Y & Y & Y & Y & Y \\
\emph{Observability} & Y & Y & Y & Y & Y & Y & Y & Y \\
\emph{Guardrails} & P & -- & P & Y & Y & Y & P & -- \\
\textbf{Agent Contract} & -- & -- & -- & -- & -- & -- & -- & -- \\
\end{longtable}

The table reveals a consistent pattern: existing frameworks provide
operational controls but lack formal governance mechanisms. Recent
practitioner perspectives frame ``agent engineering'' as iterative
refinement for reliability \citep{langchain2025engineering}, but do not
address resource governance. The following section presents Agent
Contracts as a framework that fills this gap.

\section{The Agent Contract
Framework}\label{the-agent-contract-framework}

\subsection{Contract Definition}\label{contract-definition}

An \textbf{Agent Contract} \(C\) is defined as a seven-tuple:

\[C = (I, O, S, R, T, \Phi, \Psi)\]

The components capture the complete specification for bounded agent
execution. The input specification \(I\) defines the schema and
constraints for acceptable inputs. The output specification \(O\)
defines the schema and quality criteria for deliverables. The skill set
\(S\) enumerates the capabilities (tools, functions, and knowledge
domains) available to the agent. Resource constraints \(R\) specify a
multi-dimensional budget governing consumption. Temporal constraints
\(T\) establish time-related boundaries and duration limits. Success
criteria \(\Phi\) define measurable conditions for contract fulfillment.
Finally, termination conditions \(\Psi\) specify events that end the
contract regardless of fulfillment.

This formulation synthesizes contract theory in economics, where
contracts align incentives and define obligations between parties
\citep{bolton2005contract}, with real-time systems theory, where
correctness depends on meeting explicit timing constraints
\citep{buttazzo2011hard}. The contract serves as both specification
(defining what the agent should do) and governance mechanism
(constraining how the agent may operate).

An important distinction separates \(I\) from \(R\). The input
specification \(I\) defines \emph{what} the agent receives: task
content, context, and parameters. The resource constraints \(R\) define
\emph{how much} the agent may consume while processing: token budgets,
API call limits, time bounds. An agent may receive a small input but
consume many resources through complex reasoning, or receive a large
input but consume few resources through simple transformation.

\subsection{Contract Components}\label{contract-components}

\textbf{Input and Output Specifications.} The input specification
\(I = (\sigma_I, \mathcal{V}_I, \mathcal{P}_I)\) comprises the input
schema \(\sigma_I\), validation rules \(\mathcal{V}_I\), and
preprocessing transformations \(\mathcal{P}_I\). For example, a code
review agent might specify \(\sigma_I\) as
\texttt{\{repository:\ string,\ pr\_id:\ integer\}}, \(\mathcal{V}_I\)
as \texttt{pr\_id\ \textgreater{}\ 0}, and \(\mathcal{P}_I\) as
\texttt{fetch\_diff(pr\_id)}.

The output specification \(O = (\sigma_O, Q_{min}, \mathcal{F}_O)\)
comprises the output schema \(\sigma_O\), minimum acceptable quality
threshold \(Q_{min}\), and formatting requirements \(\mathcal{F}_O\).
For the code review agent, \(\sigma_O\) might be
\texttt{\{summary:\ string,\ issues:\ list,\ approval:\ boolean\}},
\(Q_{min} = 0.8\) (requiring 80\% of issues correctly identified), and
\(\mathcal{F}_O\) as markdown format with severity labels.

\textbf{Skills and Capabilities.} The skill set
\(S = \{s_1, s_2, ..., s_m\}\) where \(s_i \in \mathcal{S}_{available}\)
enumerates what the agent \emph{can} do \citep{zhang2025skills}. Each
skill \(s_i\) may have associated costs \(c(s_i)\) and success
probabilities \(p(s_i)\). Skills encompass tool invocations (web search,
code execution, API calls), knowledge domains (legal, medical,
technical), and cognitive capabilities (reasoning, planning,
summarization). Recent industry standards for agent skills employ
\emph{progressive disclosure}---loading metadata
(\$\sim\(50 tokens) initially and full specifications (\)\sim\$500+
tokens) on-demand \citep{anthropic2025skills}---reflecting
resource-aware design even in capability definitions. The contract
restricts the agent to skills in \(S\); for example, an agent contracted
for data analysis cannot invoke payment APIs even if technically
accessible.

\textbf{Resource Constraints.} The resource constraint
\(R = \{r_1, r_2, ..., r_n\}\) defines a multi-dimensional budget.
Common resource dimensions include:

{\def\LTcaptype{none} 
\begin{longtable}[]{@{}llll@{}}
\toprule\noalign{}
Resource & Symbol & Unit & Example \\
\midrule\noalign{}
\endhead
\bottomrule\noalign{}
\endlastfoot
LLM Tokens & \(r_{tok}\) & tokens & 100,000 \\
API Calls & \(r_{api}\) & calls & 50 \\
Iterations & \(r_{iter}\) & rounds & 10 \\
Web Searches & \(r_{web}\) & queries & 10 \\
Compute Time & \(r_{cpu}\) & seconds & 300 \\
External Cost & \(r_{cost}\) & USD & 5.00 \\
\end{longtable}
}

Each resource \(r_i\) has a budget \(b_i\) and consumption function
\(c_i(t)\). Constraint satisfaction requires
\(\forall i: c_i(t) \leq b_i\).

\textbf{Temporal Constraints.} The temporal constraint
\(T = (t_{start}, \tau)\) comprises the contract activation timestamp
\(t_{start}\) and the contract duration (time-to-live) \(\tau\). The
constraint requires \(t_{current} - t_{start} \leq \tau\). While users
often think in terms of deadlines (``complete by 5pm''), duration is the
operational primitive---a deadline is simply
\(t_{deadline} = t_{start} + \tau\).

\textbf{Success Criteria and Termination.} Success criteria
\(\Phi = \{(\phi_1, w_1), (\phi_2, w_2), ..., (\phi_k, w_k)\}\) pair
measurable conditions \(\phi_i\) with weights \(w_i\). The contract is
fulfilled when \(\sum w_i \cdot \mathbb{1}[\phi_i] \geq \theta\) for
threshold \(\theta\). Conditions may include task completion
(\texttt{all\_items\_processed}), quality metrics
(\texttt{accuracy\ \textgreater{}\ 0.95}), or business logic
(\texttt{response\_generated\ AND\ reviewed}).

The output quality threshold \(Q_{min}\) (in specification \(O\)) is a
\emph{structural requirement} on output format and minimum
acceptability. Success criteria \(\Phi\) are \emph{fulfillment
conditions} that may include quality checks (e.g.,
\(\phi_1 = Q \geq Q_{min}\)) along with other conditions. An agent might
produce output meeting \(Q_{min}\) but fail \(\Phi\) if other criteria
are unmet.

Termination conditions
\(\Psi = \{\psi_1 \lor \psi_2 \lor ... \lor \psi_l\}\) define when the
contract ends regardless of success. Common termination conditions
include resource exhaustion (\(\exists r_i: c_i \geq b_i\)), duration
expiration (\(t - t_{start} > \tau\)), explicit cancellation (external
signal), and unrecoverable errors (critical failure states). The
existential quantifier ensures that exceeding \emph{any} resource budget
causes termination, not just aggregate overruns.

\subsection{Contract Lifecycle}\label{contract-lifecycle}

Contracts progress through distinct states following the transition
pattern DRAFTED → ACTIVE → \{FULFILLED, VIOLATED, EXPIRED, TERMINATED\}.
A contract begins in the DRAFTED state, where its parameters are
specified but execution has not begun. Activation transitions the
contract to ACTIVE, at which point resources are reserved and monitoring
begins. From ACTIVE, the contract reaches exactly one of four terminal
states.

The terminal state FULFILLED indicates that success criteria \(\Phi\)
were satisfied within all resource and temporal constraints. VIOLATED
indicates that some constraint in \(R\) or \(T\) was breached before
success criteria were met. EXPIRED indicates that the duration \(\tau\)
was exceeded. TERMINATED indicates external cancellation, regardless of
progress toward success criteria.

Transitions between states are governed by formal guard conditions:

{\def\LTcaptype{none} 
\begin{longtable}[]{@{}
  >{\raggedright\arraybackslash}p{(\linewidth - 4\tabcolsep) * \real{0.2143}}
  >{\raggedright\arraybackslash}p{(\linewidth - 4\tabcolsep) * \real{0.1786}}
  >{\raggedright\arraybackslash}p{(\linewidth - 4\tabcolsep) * \real{0.6071}}@{}}
\toprule\noalign{}
\begin{minipage}[b]{\linewidth}\raggedright
From
\end{minipage} & \begin{minipage}[b]{\linewidth}\raggedright
To
\end{minipage} & \begin{minipage}[b]{\linewidth}\raggedright
Guard Condition
\end{minipage} \\
\midrule\noalign{}
\endhead
\bottomrule\noalign{}
\endlastfoot
DRAFTED & ACTIVE &
\(\text{activate}() \land \text{resources\_available}()\) \\
ACTIVE & FULFILLED &
\(\sum w_i \cdot \mathbb{1}[\phi_i] \geq \theta\) \\
ACTIVE & VIOLATED & \(\exists r_i: c_i \geq b_i\) \\
ACTIVE & EXPIRED & \(t - t_{start} > \tau\) \\
ACTIVE & TERMINATED & \(\text{cancel\_signal}()\) \\
\end{longtable}
}

The lifecycle model ensures clear accountability. Every contract reaches
exactly one terminal state, enabling unambiguous resource release and
audit logging.

\section{Resource Tracking and
Monitoring}\label{resource-tracking-and-monitoring}

The preceding chapter defined what contracts \emph{are}: their
structure, components, and lifecycle. This chapter addresses how
resources are \emph{tracked} during execution. While the framework
tracks multiple resource types (tokens, API calls, tool invocations,
compute time, cost), we focus here on two foundational aspects: how
token budgets decompose into measurable categories, and how runtime
monitoring provides visibility into constraint utilization.

\subsection{Token Budget
Decomposition}\label{token-budget-decomposition}

Modern LLMs distinguish input, reasoning, and output tokens. We model
this as \(R_{tok} = (r_{in}, r_r, r_{out})\). Since input tokens
\(r_{in}\) are largely determined by task context, the
\emph{controllable budget} is \(B_{ctrl} = B_{tok} - r_{in}\),
representing tokens available for reasoning and output. This
decomposition enables fine-grained monitoring across categories,
supporting both real-time adaptation and post-hoc analysis.

\subsection{Runtime Monitoring}\label{runtime-monitoring}

During execution, a monitoring system tracks both resource consumption
and temporal progress in real-time. The monitor function
\(\text{Monitor}: (C, t) \rightarrow
(\vec{c}, \vec{u}, \tau_{util})\) takes a contract \(C\) and current
time \(t\) and returns three values: the resource consumption vector
\(\vec{c}\), the resource utilization vector
\(\vec{u} = \vec{c} / \vec{b}\) (computed element-wise), and the
duration utilization \(\tau_{util} = (t - t_{start})/\tau\) ranging from
0 to 1. The agent can query these values at any time to adapt its
strategy as constraints tighten. Note that utilization is monotonically
non-decreasing since resource consumption is cumulative.

A useful aggregate metric captures the most-constrained resource:

\[\text{utilization}(t) = \max\left(\frac{t_{current} - t_{start}}{\tau}, \max_i \frac{c_i(t)}{b_i}\right)\]

This single value summarizes how close the agent is to any constraint
boundary, enabling simple threshold-based policies (e.g., ``warn when
utilization exceeds 80\%'') without requiring sophisticated
optimization.

\textbf{Communicating Budget to Agents.} A current approach is
\emph{budget-aware prompting}: injecting remaining budget into system
prompts or providing dynamic status updates during execution. This
enables agents to self-regulate---producing concise outputs when budget
is tight or taking exploratory actions when resources are ample. As the
field matures, native solutions may emerge (e.g., models with built-in
resource awareness), but prompt-based communication provides a practical
mechanism with current infrastructure.

\section{Multi-Agent Coordination Under
Contracts}\label{multi-agent-coordination-under-contracts}

The preceding sections establish contracts for individual agents.
However, complex tasks often require multiple agents working together: a
researcher gathering data, an analyzer identifying patterns, a reporter
synthesizing findings. This extension from single-agent to multi-agent
governance raises new questions: How should a parent budget be divided
among child agents? What happens when one agent exceeds its allocation
while others remain under budget? How can the system guarantee that
aggregate consumption respects the original constraint?

These questions have theoretical grounding in coordination theory.
Malone and Crowston (1994) \citep{malone1994interdisciplinary}
identified managing shared resources as a fundamental coordination
problem. In LLM-based multi-agent systems, the shared resource is
typically the token budget (and associated cost), but the same
principles apply to API call limits, compute time, and other constrained
resources. The key insight is that contracts provide a natural unit of
delegation. When an orchestrator creates subcontracts for workers, the
contract specification ensures that each worker operates within defined
bounds, and the aggregate of those bounds respects the parent
constraint.

\subsection{Conservation Laws and Budget
Allocation}\label{conservation-laws-and-budget-allocation}

When multiple agents collaborate, contracts must govern how resources
flow between them. The fundamental constraint is \textbf{conservation}:
total consumption cannot exceed the system budget:

\[\sum_{j \in \text{agents}} c_j^{(r)} \leq B^{(r)} \quad \forall r \in R\]

This invariant holds regardless of execution pattern---sequential,
parallel, hierarchical, or competitive.

\textbf{Initial Allocation.} Before execution, the total budget \(B\)
must be divided among agents. Three strategies apply depending on
available information:

\begin{itemize}
\tightlist
\item
  \emph{Proportional}:
  \(b_j = \omega_j / \sum \omega_k \cdot (B - B_{reserve})\), where
  \(\omega_j\) reflects estimated task complexity
\item
  \emph{Equal}: \(b_j = (B - B_{reserve})/n\), when complexity is
  unknown
\item
  \emph{Negotiated}: Agents request budgets; a coordinator allocates
  based on requests with caps to prevent over-claiming
\end{itemize}

A reserve buffer \(B_{reserve}\) (typically 10--15\%) accommodates
coordination overhead and unexpected costs.

\textbf{Dynamic Reallocation.} As agents complete, unused budget returns
to a shared pool:

\[B_{available}(t) = B_{reserve} + \sum_{j \in \text{completed}} (b_j - c_j)\]

This enables \emph{budget pooling}---efficient agents effectively
subsidize those requiring more resources, improving overall throughput
while maintaining total budget discipline.

\subsection{Coordination Patterns Through a Contract
Lens}\label{coordination-patterns-through-a-contract-lens}

Recent work has identified recurring design patterns for agentic AI
systems, including routing, orchestration, parallelization, and
iterative refinement
\citep{ng2024agentic, anthropic2024agents, google2025patterns}.

We organize these patterns through a \textbf{contract-centric lens},
focusing on how resource constraints govern each pattern's behavior and
where contracts provide critical safety guarantees. We focus here on two
typical \textbf{control flow patterns}, task routing and delegation
decisions, since these are where contracts add the most value. Execution
within any pattern can be sequential or parallel; these are orthogonal
concerns that compose naturally (as we discuss below).

\textbf{Routing.} An input classifier directs requests to specialized
handlers based on task characteristics, selecting the best-suited agent
for each input and allocating the corresponding budget
\citep{anthropic2024agents}. Budget is reserved per potential branch;
unused allocations return to the pool. This enables separation of
concerns--- specialized agents outperform generalists on their specific
tasks.

When specialists have explicit contracts, routing becomes more
principled: the router matches task requirements against specialist
capabilities, resource profiles, and success criteria. Furthermore, with
well-defined contracts, the router need not be limited to a fixed pool;
it can dynamically instantiate or configure an agent specifically for
the required contract. This blurs the line between routing and
orchestration, with the contract serving as the specification for agent
creation, not just agent selection.

\textbf{Orchestrator-Workers.} A central orchestrator dynamically
decomposes tasks and delegates to worker agents, synthesizing results
\citep{anthropic2024agents, google2025patterns}. This pattern can extend
to multiple levels (hierarchical orchestration).

From a contract perspective, this pattern is particularly significant:
the orchestrator \textbf{drafts and issues subcontracts} to workers.
Each subcontract specifies the worker's task, allocated budget, and
success criteria:

\[\text{orchestrator}(C_{parent}) \rightarrow \{C_i = (I_i, O_i, S_i, R_i, T_i, \Phi_i, \Psi_i)\}_{i=1}^{k}\]

This frames \textbf{contracting as a capability}: the orchestrator must
understand the contract framework to effectively delegate work. The
conservation law (Section 6.1) constrains subcontract allocation:
\(\sum R_i \leq R_{parent}\).

The implications of contracting as a capability are significant:

{\def\LTcaptype{none} 
\begin{longtable}[]{@{}
  >{\raggedright\arraybackslash}p{(\linewidth - 2\tabcolsep) * \real{0.5000}}
  >{\raggedright\arraybackslash}p{(\linewidth - 2\tabcolsep) * \real{0.5000}}@{}}
\toprule\noalign{}
\begin{minipage}[b]{\linewidth}\raggedright
Implication
\end{minipage} & \begin{minipage}[b]{\linewidth}\raggedright
Description
\end{minipage} \\
\midrule\noalign{}
\endhead
\bottomrule\noalign{}
\endlastfoot
\emph{Recursive delegation} & Agents spawn sub-agents that themselves
have contracting capability, enabling hierarchical self-organization \\
\emph{Bounded autonomy} & Even highly capable orchestrators remain
governed by their parent contract: they can create subcontracts but
cannot exceed their own constraints \\
\emph{Dynamic team formation} & Agents form coalitions and delegate work
without centralized coordination, as long as conservation laws are
satisfied \\
\emph{Dynamic agent instantiation} & Rather than selecting from a fixed
pool, agents instantiate specialists on-demand; the contract becomes the
specification for agent creation, not just selection \\
\emph{Meta-governance} & Contracts can govern the creation of other
contracts, enabling principled scaling of multi-agent systems \\
\end{longtable}
}

These patterns demonstrate how contracts can govern increasingly complex
multi-agent systems. However, practical enforcement faces fundamental
constraints that shape what contracts can and cannot guarantee.

\section{Fundamental Limitations and Practical
Enforcement}\label{fundamental-limitations-and-practical-enforcement}

\subsection{Single-Call Enforcement
Constraints}\label{single-call-enforcement-constraints}

A critical limitation exists. Token consumption is only known
\emph{after} an LLM call completes, not during execution:

\[c_{tok}(t) = \begin{cases}
\text{unknown} & \text{during API call} \\
c_{actual} & \text{after API returns}
\end{cases}\]

Even streaming APIs provide total usage metadata only after generation
completes. This asymmetry has three important consequences: contracts
cannot prevent a single expensive call from exceeding budget, contracts
can prevent subsequent calls after budget is exceeded, and the primary
value is therefore multi-call protection.

\subsection{Enforcement Capabilities}\label{enforcement-capabilities}

Agent Contracts employ two complementary enforcement layers. \emph{Soft
enforcement} operates cooperatively: budget-aware prompts (Section 5)
communicate remaining resources, enabling agents to self-regulate. This
is best-effort---agents may ignore or exceed stated constraints, as
documented by the ``token elasticity'' phenomenon
\citep{han2024tokenbudget}. \emph{Hard enforcement} operates
structurally: an external monitor tracks consumption after each action
and halts execution when constraints are breached, regardless of agent
behavior. This follows the \emph{agent harness} pattern emerging in
production systems---an infrastructure layer that wraps the agent,
intercepts its actions, and enforces invariants without relying on the
agent's cooperation. Hard enforcement requires no model-level support;
it operates at the orchestration layer between actions. This structural
approach replaces the normative sanctions of social commitments (Section
2): where persistent agents are sanctioned for violations, ephemeral LLM
agents are simply halted, with accountability attributed to the
orchestrator's allocation strategy and recorded for audit.

Despite single-call limitations, hard enforcement provides substantial
value across multiple enforcement dimensions:

{\def\LTcaptype{none} 
\begin{longtable}[]{@{}
  >{\raggedright\arraybackslash}p{(\linewidth - 4\tabcolsep) * \real{0.2800}}
  >{\centering\arraybackslash}p{(\linewidth - 4\tabcolsep) * \real{0.1000}}
  >{\raggedright\arraybackslash}p{(\linewidth - 4\tabcolsep) * \real{0.6200}}@{}}
\toprule\noalign{}
\begin{minipage}[b]{\linewidth}\raggedright
Constraint
\end{minipage} & \begin{minipage}[b]{\linewidth}\centering
Enf.
\end{minipage} & \begin{minipage}[b]{\linewidth}\raggedright
Mechanism
\end{minipage} \\
\midrule\noalign{}
\endhead
\bottomrule\noalign{}
\endlastfoot
Multi-call budgets & Full & Stop after cumulative threshold \\
Iteration limits & Full & Count and halt at \(r_{iter}\) \\
API call limits & Full & Count external invocations \\
Duration limits & Full & Check elapsed time before each action \\
Cost ceilings & Approx & Track cumulative cost; bound via
\texttt{max\_tokens} \\
\end{longtable}
}

Contracts provide particularly high value in five scenarios. First,
retry loops benefit from contracts that prevent runaway costs when tasks
fail repeatedly. Second, validation cycles benefit from contracts that
limit iterative refinement (e.g., code generation → test → fix loops).
Third, multi-agent workflows benefit from contracts that stop downstream
agents when upstream exceeds budget. Fourth, tool-heavy agents benefit
from contracts that control cumulative cost of web searches and API
calls. Fifth, long-running sessions benefit from contracts that enforce
session-level budgets across many interactions.

\subsection{Future Infrastructure
Requirements}\label{future-infrastructure-requirements}

True real-time enforcement would require further API-level changes.
While some providers now offer cumulative token counts during streaming,
key capabilities remain missing: interruptible generation allowing
mid-generation cancellation with partial output, token reservation to
pre-allocate budget with guaranteed hard limits, and budget-aware
inference enabling models to respect token budgets natively, as explored
in recent work on budget-aware reasoning
\citep{han2024tokenbudget, liu2025budgetthinker}.

Such infrastructure would extend hard enforcement from the inter-call
level (where it operates today) to the intra-call level, enabling
real-time budget guarantees within a single generation. Until then,
Agent Contracts provide hard enforcement between actions and soft
enforcement within them.

\section{Empirical Evaluation}\label{sec-evaluation}

We validate Agent Contracts through four complementary experiments
spanning single-agent and multi-agent settings. Each experiment targets
specific framework components; collectively they validate the
enforcement mechanisms (\(R\), \(T\), \(\Phi\), \(\Psi\)) and
coordination primitives (conservation laws, contract delegation) that
distinguish Agent Contracts from existing approaches. Experiments use a
reference implementation with Google ADK (Code Review, Research
Pipeline, Crisis Communication) or LiteLLM (Strategy Modes), using
Gemini 2.5 Flash and Flash-Lite (knowledge cutoff: January
2025).\footnote{Implementation available at
  https://github.com/flyersworder/agent-contracts. The framework is
  under active development; we provide experiment code and data for
  reproducibility.} Statistical analysis employs bootstrap confidence
intervals (10,000 resamples) using the percentile method.

\subsection{Experimental Overview}\label{experimental-overview}

The following table summarizes our experimental design, with each
experiment validating a distinct aspect of the framework.

{\def\LTcaptype{none} 
\begin{longtable}[]{@{}
  >{\raggedright\arraybackslash}p{(\linewidth - 6\tabcolsep) * \real{0.2000}}
  >{\raggedleft\arraybackslash}p{(\linewidth - 6\tabcolsep) * \real{0.1200}}
  >{\raggedright\arraybackslash}p{(\linewidth - 6\tabcolsep) * \real{0.3400}}
  >{\raggedright\arraybackslash}p{(\linewidth - 6\tabcolsep) * \real{0.3400}}@{}}
\toprule\noalign{}
\begin{minipage}[b]{\linewidth}\raggedright
Experiment
\end{minipage} & \begin{minipage}[b]{\linewidth}\raggedleft
n
\end{minipage} & \begin{minipage}[b]{\linewidth}\raggedright
Validates
\end{minipage} & \begin{minipage}[b]{\linewidth}\raggedright
Key Result
\end{minipage} \\
\midrule\noalign{}
\endhead
\bottomrule\noalign{}
\endlastfoot
Code Review & 70 & Runaway prevention & 90\% token reduction \\
Research Pipeline & 50 & Conservation laws & 0 violations \\
Strategy Modes & 50 & Satisficing tradeoffs & 70\%→86\% success \\
Crisis Comm. & 24 & Failure prevention & 23\% fewer tokens \\
\end{longtable}
}

\subsection{Runaway Prevention in Iterative
Workflows}\label{runaway-prevention-in-iterative-workflows}

The ``\$47K problem'' from the Introduction illustrates the danger of
unbounded agent loops. We evaluate iteration governance using a
Coder↔Reviewer pipeline where agents iteratively refine code solutions.
The Coder writes Python code; the Reviewer executes it against test
cases using a \texttt{test\_code} tool and either approves or requests
revision. The experiment uses 70 problems from LiveCodeBench
\citep{jain2024livecodebench} (released post-February 2025, after model
cutoff): 31 easy and 39 medium difficulty.

\textbf{Design.} Each problem is executed twice in a within-subjects
design: CONTRACTED (50K token budget, max 3 iterations) versus
UNCONTRACTED (no token limits, max 6 iterations). CONTRACTED agents
receive budget-aware prompts and dynamic status updates showing both
token consumption and iteration progress; UNCONTRACTED agents use
identical base prompts without resource awareness.

\textbf{Results.} CONTRACTED execution achieves 90\% token reduction
compared to UNCONTRACTED (\(p=0.0007\), paired \(t\)-test), with 525×
lower variance (5.29B vs 10.1M)---directly addressing the ``\$47K
problem.'' The success rate difference of 7.1 percentage points (52.9\%
vs 60.0\%) is not statistically significant (\(p=0.13\)). Governance
value increases with task complexity: medium-difficulty problems show
92\% token savings versus 76\% for easy.

{\def\LTcaptype{none} 
\begin{longtable}[]{@{}lrrrr@{}}
\toprule\noalign{}
Metric & UNCONTRACTED & CONTRACTED & Change & \(p\)-value \\
\midrule\noalign{}
\endhead
\bottomrule\noalign{}
\endlastfoot
Token Usage & 34,606 & 3,461 & −90\% & 0.0007*** \\
Variance & 5.29B & 10.1M & 525× lower & --- \\
Iterations & 3.00 & 1.71 & −43\% & \textless0.0001*** \\
LLM Calls & 9.0 & 4.5 & −50\% & \textless0.0001*** \\
Success Rate & 60.0\% & 52.9\% & −7.1pp & 0.13 (NS) \\
\end{longtable}
}

A complementary single-agent experiment with 24 time-critical crisis
communication scenarios provides additional evidence: agents with
explicit quality thresholds (\(Q \geq 0.80\)) and iteration limits
achieved 23\% token reduction (\(p=0.005\)) with statistically
equivalent quality (\(p=0.32\)). Notably, one UNCONTRACTED agent failed
entirely---stuck in an evaluation loop without submitting output---while
the CONTRACTED version succeeded, demonstrating that iteration
governance prevents agent failures, not just improves efficiency.

\subsection{Conservation Laws in Multi-Agent
Coordination}\label{conservation-laws-in-multi-agent-coordination}

We evaluate conservation laws and contract delegation (Section 6) using
a three-agent research pipeline: Researcher → Analyzer → Reporter. The
orchestrator delegates sub-contracts to each worker via
\texttt{DelegatingAdkAgent}, enforcing both the conservation invariant
\(\sum b_i \leq B\) and per-tool limits at allocation time.

\textbf{Design.} Fifty research topics across five categories
(technology, science, business, health, society) are evaluated.
CONTRACTED agents receive explicit budget allocations with per-tool
limits (e.g., Researcher limited to 6 web searches) and budget-aware
prompts; UNCONTRACTED agents operate without constraints. Quality is
assessed via multi-judge LLM evaluation following best practices for
rating indeterminacy \citep{guerdan2025indeterminacy}.

\textbf{Results.} CONTRACTED execution achieves zero conservation
violations across all 50 trials, demonstrating that conservation laws
(\(\sum b_i \leq B\)) are fully enforceable. In one trial, the
enforcement mechanism successfully detected and halted a runaway agent
that exceeded its 40K token budget (56K consumed), providing evidence
that runtime enforcement works as designed. Quality variance is 26.7×
lower than UNCONTRACTED (\(\sigma\): 1.75 vs 9.07), though partly driven
by one catastrophic failure in the UNCONTRACTED condition. Excluding
this outlier, CONTRACTED still shows 1.4× lower variance (88.5\%
Bayesian probability). The key insight is not mean quality improvement
but variance reduction---contracts eliminate catastrophic failures where
agents consume resources without producing useful output.

\subsection{Quality-Resource Tradeoffs via Contract
Modes}\label{quality-resource-tradeoffs-via-contract-modes}

Contract modes operationalize Simon's satisficing principle
\citep{simon1955behavioral}: agents achieve acceptable quality within
defined resource bounds. We test whether different contract
configurations produce measurable behavioral differences using logic
reasoning problems from OpenR1 (released February 2025, after model
cutoff).

\textbf{Design.} Fifty medium-difficulty logic puzzles are evaluated
under three contract modes: URGENT (no extended reasoning, 30s timeout),
ECONOMICAL (low reasoning effort, 60s timeout), and BALANCED (medium
reasoning effort, 90s timeout). Success requires exact numeric answer
match.

\textbf{Results.} Contract modes produce a clear quality-resource
gradient. BALANCED mode achieves 86\% success rate versus 70\% for
URGENT, investing 75\% more tokens for 16 percentage points higher
success (\(p \approx 0.05\)). The \texttt{reasoning\_effort} parameter
provides direct control: users requiring speed accept lower accuracy
(URGENT); those requiring accuracy invest more resources (BALANCED).

{\def\LTcaptype{none} 
\begin{longtable}[]{@{}lrrrr@{}}
\toprule\noalign{}
Mode & Success Rate & Reasoning Tokens & Avg Time & Timeout Rate \\
\midrule\noalign{}
\endhead
\bottomrule\noalign{}
\endlastfoot
URGENT & 70\% & 0 & 6.9s & 26\% \\
ECONOMICAL & 76\% & 718 & 12.5s & 14\% \\
BALANCED & 86\% & 1,519 & 16.9s & 10\% \\
\end{longtable}
}

Across all four experiments, the consistent finding is not cost
optimization but \emph{governance}: contracts transform unpredictable
agent behavior into bounded, auditable operations. Code Review
demonstrates multi-dimensional resource enforcement (90\% token
reduction, 525× variance reduction). Research Pipeline validates
conservation laws and runtime enforcement (zero conservation violations;
one runaway agent detected and halted). Crisis Communication shows that
quality thresholds prevent failures, not just reduce costs. Strategy
Modes confirms that contract configurations operationalize satisficing
tradeoffs (70\%→86\% success).

{\def\LTcaptype{none} 
\begin{longtable}[]{@{}
  >{\raggedright\arraybackslash}p{(\linewidth - 6\tabcolsep) * \real{0.2200}}
  >{\raggedright\arraybackslash}p{(\linewidth - 6\tabcolsep) * \real{0.1600}}
  >{\raggedright\arraybackslash}p{(\linewidth - 6\tabcolsep) * \real{0.3200}}
  >{\raggedright\arraybackslash}p{(\linewidth - 6\tabcolsep) * \real{0.3000}}@{}}
\toprule\noalign{}
\begin{minipage}[b]{\linewidth}\raggedright
Framework Component
\end{minipage} & \begin{minipage}[b]{\linewidth}\raggedright
Experiment
\end{minipage} & \begin{minipage}[b]{\linewidth}\raggedright
Validation
\end{minipage} & \begin{minipage}[b]{\linewidth}\raggedright
Specification
\end{minipage} \\
\midrule\noalign{}
\endhead
\bottomrule\noalign{}
\endlastfoot
Budget awareness & All & Prompts; dynamic status &
\texttt{"Budget:\ \{used\}/\{total\}"} \\
Resource constraints \(R\) & Code Review & Tokens, iterations, calls &
\(r_{tok}\)=50K, \(r_{iter}\)=3, \(r_{llm}\)=6 \\
Agent delegation & Research & Conservation, per-tool limits &
\(\sum R_i \leq 100K\); \texttt{web\_search}\(\leq 6\) \\
Success criteria \(\Phi\) & Crisis Comm. & Quality thresholds &
\(Q_{min} = 0.80\) \\
Contract modes & Strategy & URGENT/ECON./BALANCED & \(\tau\): 30s / 60s
/ 90s \\
\end{longtable}
}

\section{Conclusion}\label{conclusion}

Agent Contracts provide a formal framework for governing autonomous AI
agents through explicit resource and temporal constraints. The contract
specification \(C = (I, O, S, R, T, \Phi, \Psi)\) unifies resource,
temporal, and quality governance with conservation laws for multi-agent
delegation.

Empirical validation demonstrates that the formal framework translates
into practical governance: resource constraints enable 90\% token
reduction with 525× lower variance; conservation laws achieve 100\%
compliance in multi-agent delegation; and contract modes operationalize
satisficing tradeoffs (70\%→86\% success).

Several extensions merit investigation: \emph{runtime cancellation}
would require API-level support from model providers to halt
mid-execution; \emph{learning-based contract design} would enable agents
to predict budgets and draft subcontracts
\citep{han2024tokenbudget, liu2025budgetthinker}; \emph{formal
verification} of contract properties could draw on existing work on
resource-bounded MAS verification
\citep{alechina2010resource, bulling2016norm}, extending these
techniques from the governance domain we address here toward automated
reasoning about contract satisfiability and optimal budget derivation;
\emph{human-in-the-loop contracts} would specify when approval is
required; and \emph{milestone-based governance} would support continuous
agent operation. As agentic AI moves to production, formal governance
becomes essential---not only for cost control but for accountability and
value alignment, ensuring agents operate within bounds acceptable to
stakeholders. Agent Contracts provide one such foundation.

\section{References}\label{references}

\renewcommand{\bibsection}{}
\bibliography{references.bib}

\end{document}